\pdfoutput=1
\documentclass[fleqn,usenatbib]{mnras}
\usepackage{graphicx,times,epsf}             
\usepackage{longtable}
\usepackage{natbib}
\usepackage{amssymb}
\usepackage{amsmath}
\usepackage{psfig}
\usepackage{txfonts}
\usepackage[T1]{fontenc}
\usepackage[T1]{fontenc}
\DeclareRobustCommand{\VAN}[3]{#2}
\let\VANthebibliography\thebibliography
\def\thebibliography{\DeclareRobustCommand{\VAN}[3]{##3}\VANthebibliography}
\begin{document}

\title[Virgo Galaxy Variability]{Galaxy Optical Variability of Virgo Cluster: New Tracer for Environmental Influences on Galaxies}

\author[Yang et al.]{Fan Yang\href{https://orcid.org/0000-0002-6039-8212}{\includegraphics[scale=0.1]{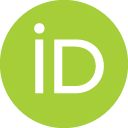}},$^{1,}$$^{2,}$$^{3}$ Richard J. Long,$^{4,}$$^{1,}$$^{5}$ Su-Su Shan,$^{1,}$$^{3}$\footnotemark[2] Jun-Qiang Ge,$^{1,}$ Rui Guo,$^{1,}$$^{3}$ \newauthor
Bo Zhang,$^{1,}$$^{3}$ Jing-Hua Gao,$^{1,}$$^{3}$ Xiang Ji,$^{1,}$$^{3}$ And Ji-Feng Liu$^{1,}$$^{3}$\footnotemark[1]\\
$^{1}$National Astronomical Observatories, Chinese Academy of Sciences, 20A
     Datun Road, Chaoyang District, Beijing 100101, China\\
     $^{2}$IPAC, Caltech, KS 314-6, Pasadena, CA 91125, USA\\
     $^{3}$School of Astronomy and Space Science, University of Chinese Academy of Sciences,
Beijing 100049, China\\
$^{4}$Department of Astronomy, Tsinghua University, Beijing 100084, China\\
$^{5}$Jodrell Bank Centre for Astrophysics, Department of Physics and Astronomy, The University of Manchester, Oxford Road, Manchester M13 9PL, UK\\
}

\label{firstpage}
\pagerange{\pageref{firstpage}--\pageref{lastpage}}
\renewcommand{\thefootnote}{\fnsymbol{footnote}}

\maketitle

\begin{abstract}
We investigate the relationship between the optical variability of galaxies and their distances from the centre of the Virgo Cluster using Palomar Transient Factory data. We define the ratio between the standard deviation of the galaxy brightness and the mean value of the standard deviation as a measure of a galaxy's optical variability. A sample of 814 Virgo galaxies with 230263 observations shows a monotonically decreasing trend of optical variability with increasing clustercentric distance. The variability level inside the cluster is 3.2$\sigma$ higher than the level outside. We fit the variability with a linear function and find that the data reject a distance-independent model. We examine 217 background galaxies for comparison and find no significant trend in galaxy variability. We assess the relation with Monte Carlo simulation by rebuilding the brightness of each galaxy. The simulation shows a monotonically decreasing relation for member galaxy variability and a distance-independent relation for background galaxies. Our result is consistent with the theory that the cold gas flowing inwards the cluster centre fuels AGN activity. This work is a new implementation of the method using optical variability to investigate the relation between galaxies evolution and their environment. 

\end{abstract}
\footnotetext[1]{Email:jfliu@nao.cas.cn}
\footnotetext[2]{Email:shansusu@nao.cas.cn}
\section{introduction}
Member galaxies in condensed cluster environments yield clues to connect the evolution of the galaxies with their surroundings. Commonly, galaxies flow towards the gravitational well in the centre of clusters. Meanwhile, gas in the galaxies is stripped by ram pressure \citep{Gunn1972}. Cluster galaxies, especially in the core of the cluster, tend to be more red and elliptical \citep[morphology-density relation;][]{Dressler1980}. Some galaxy clusters behave abnormally and have reduced temperature near the cluster centre (cool core cluster). Some clusters even present an increasing amount of cold gas in the vicinity of the cluster centre (cooling flow cluster) which can trigger star formation in member galaxies \citep[see][]{Fabian1994, Peterson2003}. The cooling of the gas in cool core and cooling flow clusters is counteracted by heating from the central active galactic nucleus(AGN) \citep{Hillel2014, Fabian2012, McDonald2015} which weakens the influence of cluster cooling flows. The precise interaction between hot gas ($\sim 10^{7}$K), cold star-forming gas (10-100K) and galaxy evolution is still subject to debate.

Some observations of star-forming galaxies support unique processes in galaxy clusters and reveal different statuses of hot and cold gas inside galaxy clusters. The star formation rate (SFR) of galaxies is reported to decline towards the cluster centre \citep{Pasquali2009, Linden2010, Rodr2019}. However, the star formation rate is reportedly not environment-dependent for actively star-forming/blue galaxies \citep{Woo2013, Wijesinghe2012}. One explanation for this scenario is that the environment takes considerable time (several Gyrs) to affect the satellite star formation rate. In addition, the time scales of galaxy gas stripping by ram pressure are reported to vary considerably from 10-100 Myr \citep{Gunn1972} to a few Gyr \citep{Balogh2000, Wetzel2013} which is close to the cluster crossing time and the galaxy depletion time \citep{Elmegreen2000, Tremonti2004, Leroy2008}.

Active galactic nuclei also play a major role in modulating the formation and evolution of their host galaxies by influencing the energy as well as the gas in the host galaxies \citep{Silk1998, Fabian2012}. The correlation between black hole and galaxy bulge mass is reported to occur as a result of repeated mergers \citep{Jahnke2011}. Gas gathering at the centre of a galaxy due to loss of angular momentum during mergers triggers intense starbursts and AGN activity \citep{Hernquist1989, Mihos1996}. However, AGN are also thought to quench star formation by exhausting the gas and heating cold gas, in a process known as feedback \citep{Kauffmann2004, Arora2019}. The AGN fraction in galaxy clusters as a function of distance to the cluster centre has been assessed in the last decade. Unlike the situation with star formation, the AGN relation in the literature shows diverse results \citep{Pasquali2009, Linden2010, Ehlert2015} which indicate that the influence of the environment on AGN activity might depend on, for instance, the type of cluster.

The cold gas in cool core and cooling flow clusters is reported to provide fuel to the AGN activity and to star formation, especially for radio-mode AGN \citep{Hillel2014, McDonald2015}. AGN activity and star formation are tightly linked in cool core clusters which indicates that AGN feedback is not enough to fully quench star formation due to the extra cold gas available \citep{Salom2003, Hillel2014, Li2015}.

Galaxy optical variability (if there is any) comes from the resident AGN \citep{Prakash2019, Salvato2009}.
Here we present a new tracer which is the radial distribution of galaxy variability for the Virgo cluster to show the environmental influence on galaxies. The paper is organized as follows. Section 2 describes our sample and the PTF data used. In section 3, we present the criterion for galaxy variability as well as its clustercentric distribution. Section 4 gives a summary of our results.

\section{Sample Selection and Observation}
\subsection{Virgo Cluster Member Galaxies}

The Virgo cluster, the closest large cluster to our local group, is reported as a cool core and cooling flow cluster \citep{Fabian1994, Hudson2010}. \citet{Binggeli1985} identifies 1277 member galaxies and 574 possible member galaxies for the Virgo cluster, their Virgo Cluster Catalog (VCC). Our galaxy sample comes from this catalog. Also, 245 galaxies are reported by \citet{Binggeli1985} as background galaxies and these are distributed as part of VCC. We use these galaxies as a control sample for comparison purposes in our work.

Virgo cluster is reported to be 16.5 Mpc away \citep{Mei2007}. The member galaxies from \citet{Binggeli1985} distribute within a radius of 1.6 Mpc from the cluster centre. The most distant background galaxy in our control sample is 1.4 Mpc away from the cluster centre. The largest clustercentric distances of both member and background galaxies in our sample are close to the virial radius of the Virgo cluster \citep[i.e. 1.72Mpc; ][]{Hoffman1980}.

\subsection{PTF Photemetry}

The Palomar Transient Factory is a time-domain sky survey, employing the 48 inch Samuel Oschin Telescope with an 8 square degree camera \citep{Law2009}. The exposure time for the PTF survey is 60 seconds, with limiting magnitudes of $m_{g^{'}}$ $\sim$ 21.3 and $m_{R} \sim$ 20.6. The source extraction and data reduction are performed by an automated pipeline in 
the Infrared Processing and Analysis centre \citep[IPAC; ][]{Laher2014}. In our work, we aim to use R band data due to the greater number of observations. The PTF data have been well-utilized by other authors \citep[e.g.][]{Pian2017,Kao2016}. The conditions under which we use the data are similar to those in the work of \citet{Cook2019}.

As our work is based on the estimation of galaxy optical variability, an assessment of the PTF photometry uncertainty is necessary. To this end, we choose 10$^5$ stars in the SDSS Stripe 82 Standard Catalog \citep{Ivezi2007}. The standard deviations of magnitude of these standard stars indicate the statistical error of the PTF photometry (shown in Figure 1). In the range of 14 to 20 magnitude, the standard deviation is less than 0.1 mag. 

The VCC galaxies were crossed matched with the Palomar Transient Factory (PTF) data base\footnote{\url{https://irsa.ipac.caltech.edu/cgi-bin/Gator/nph-scan?submit=Select&projshort=PTF}}. We retrieved 554 confirmed member galaxies with 169258 observations, 260 possible members with 61005 observations, and 217 background galaxies with 55921 observations. During cross-matching, we required the number of observational epochs for each source to be greater than 20, with magnitudes in the range of 14 to 18. The histogram of observation times for PTF VCC galaxies is given in Figure 2.  

\begin{figure}
\centering
\includegraphics[width=3.8in]{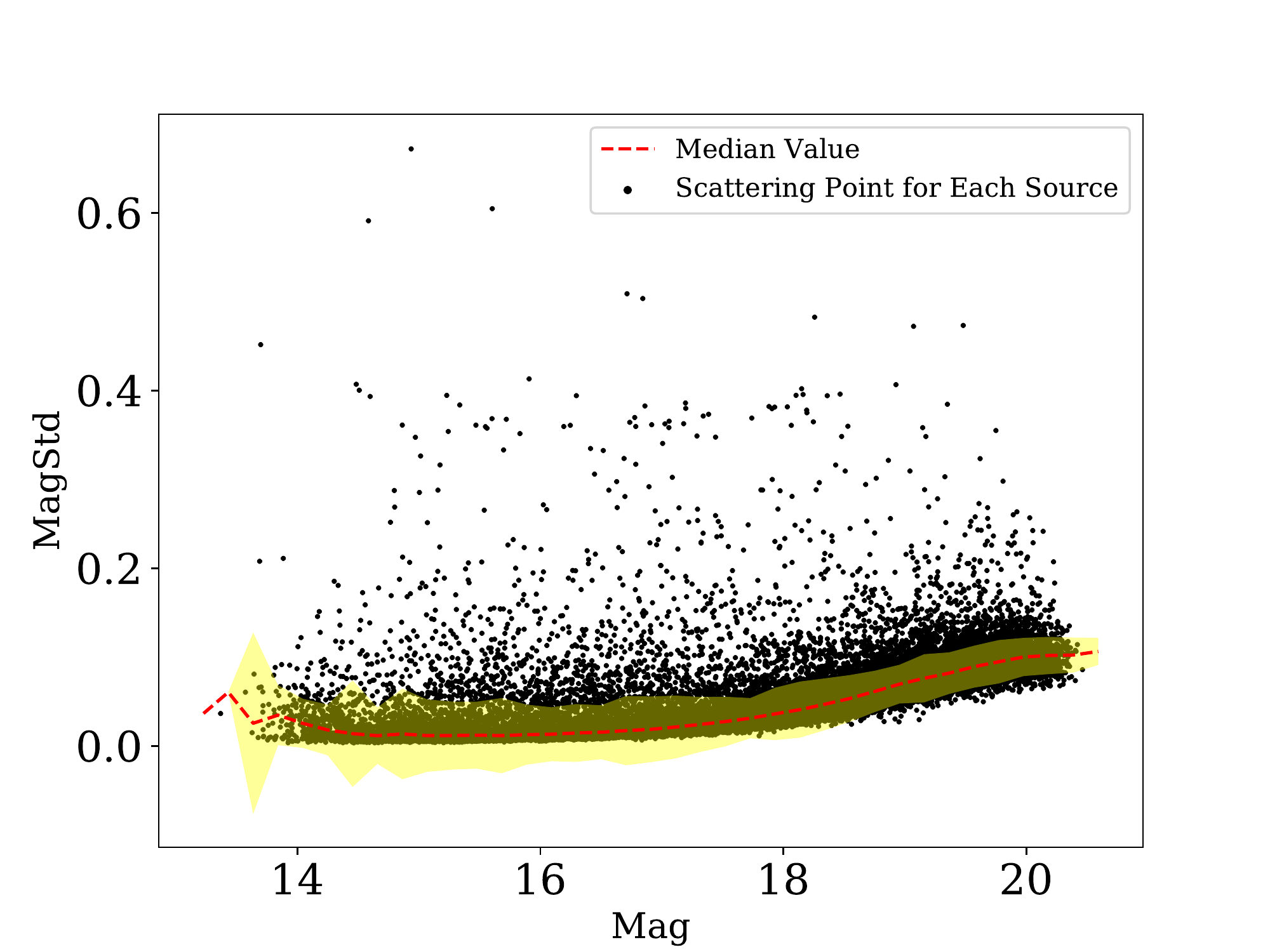}
\caption{Standard deviation of PTF standard star magnitude vs magnitude. The black points gives the scattering for each source. The red line shows the median value of the mag standard deviation. The yellow region indicates the 1$\sigma$ confidence region for the scattering.}
\end{figure}

\begin{figure}
\centering
{\includegraphics[width=3.8in]{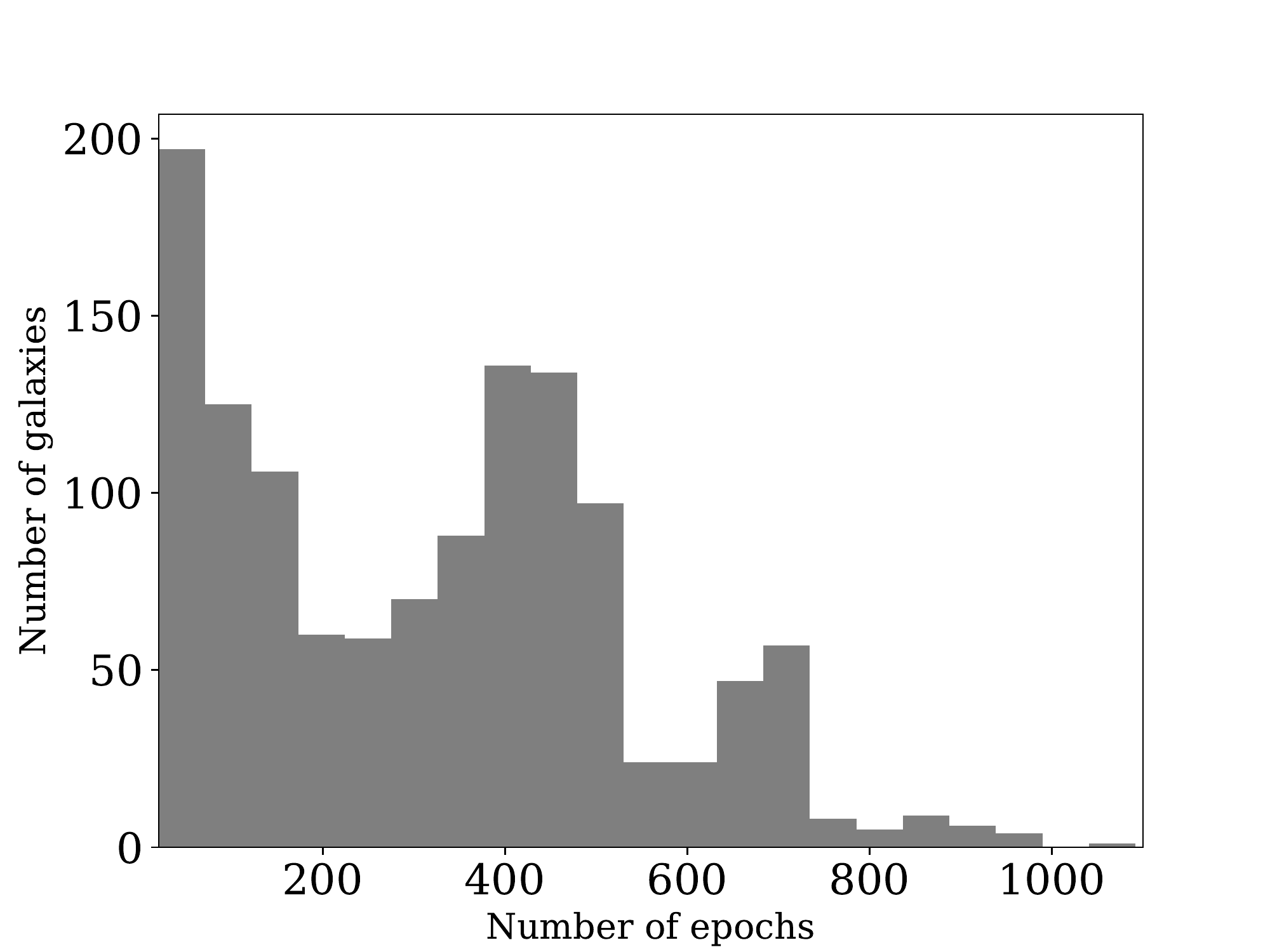}}
\caption{The number of epochs for the sample in PTF data base. The minimum number of epochs is 20, and corresponds to the start of the x-axis. }
\end{figure}

\section{Analysis and Result}
\subsection{Estimating Galaxy Optical Variability}

The optical galaxy variability comes intrinsically from the activity of galactic nuclei. Optical variability has been proven to be an important criterion for AGN detection \citep{Trevese2008,Villforth2010,Cicco2015, Prakash2019}. The standard deviation of a galaxy brightness is a combination of the galaxy intrinsic brightness variability and the observational uncertainties of the measurements. In PTF data, the observational error is connected to the brightness as shown in Figure 1. To investigate the physical properties of galaxy variability, we have to remove the influence of observational error. e.g. the uncertainty dependency upon brightness. 

Thus we define galaxy optical variability in our work as the ratio between the standard deviation of the galaxy flux among all the measurements taken of a galaxy ($\sigma_{single}$) and the mean value of the standard deviation in the flux bin enclosing the magnitude of the galaxy ($\sigma_{flux}$). Mathematically, the definition can be explained by considering the following error propagation equation.
\begin{equation}
Variability=\sigma_{single}/\sigma_{flux}=\sqrt{\frac{\sigma_{flux}^{2}\times\xi^{2}+\sigma_{galaxy}^{2}}{\sigma_{flux}^{2}}}=\sqrt{\xi^{2}+\frac{\sigma_{galaxy}^{2}}{\sigma_{flux}^{2}}}        
\end{equation}
$\sigma_{galaxy}$ indicates the galaxy intrinsic variability. $\xi$ is a random variable indicating the ratio between the standard deviation of one galaxy caused by photometry uncertainty and $\sigma_{flux}$. The flux is derived from photometry magnitude according to the Flux-Apparent Magnitude Relationship. The brightness bin here is set to be 0.2 in the unit of mag. We also performed tests with bin sizes of 0.1, 0.5 which gave negligible differences.

\subsection{Galaxy Variability and Clustercentric Distance Relation}

Using the variability estimates from the previous sections, we display the radial distribution of the galaxy variability for the Virgo Cluster in Figure 3. The plot shows a large scatter due to the combination of the diversity of galaxy variability and observational uncertainty.

The points are not uniformly distributed across the plot with the points in the upper right corner being particularly sparse. This can be explained by the fact that some galaxies in the inner part of the cluster present optical variability due to AGN. Theses sources have higher statistical variability levels and thus occupy the space in the upper left corner of the plot. 
We also give the variability of background galaxies for comparison (see Figure 3). The background galaxy variability gives a more uniform distribution.

To determine the statistical properties of the variability distribution, we calculate the mean value of the scattered points in radius bins (in Figure 3). The error bars are derived according to standard error propagation rules and are the ratio between the standard deviation of the points in the bin and the square root of the numbers of points in the bin. Any undersampling if present will be reflected in higher error bars.

The binned values indicate a monotonically decreasing variability with increasing clustercentric distance. The variability level in the cluster centre is higher than that outside at a significance level of 3.2$\sigma$. The values in all the distance bins are within 1$\sigma$ uncertainty to the fit prediction.

We fitted the binned values with a linear function as well as a flat line for comparison (in Figure 3). The best-fitting linear function gives us a reduced chi-square of 0.49 indicating that a linear fit with a negative slope is a good fit to the data, but the large uncertainties may indicate some degeneracy. The flat line fit to the data gives a reduced chi-square of 3.64 which means that a 'distance-independent' model is rejected by the data.

\begin{figure}
\centering
\includegraphics[width=3.8in]{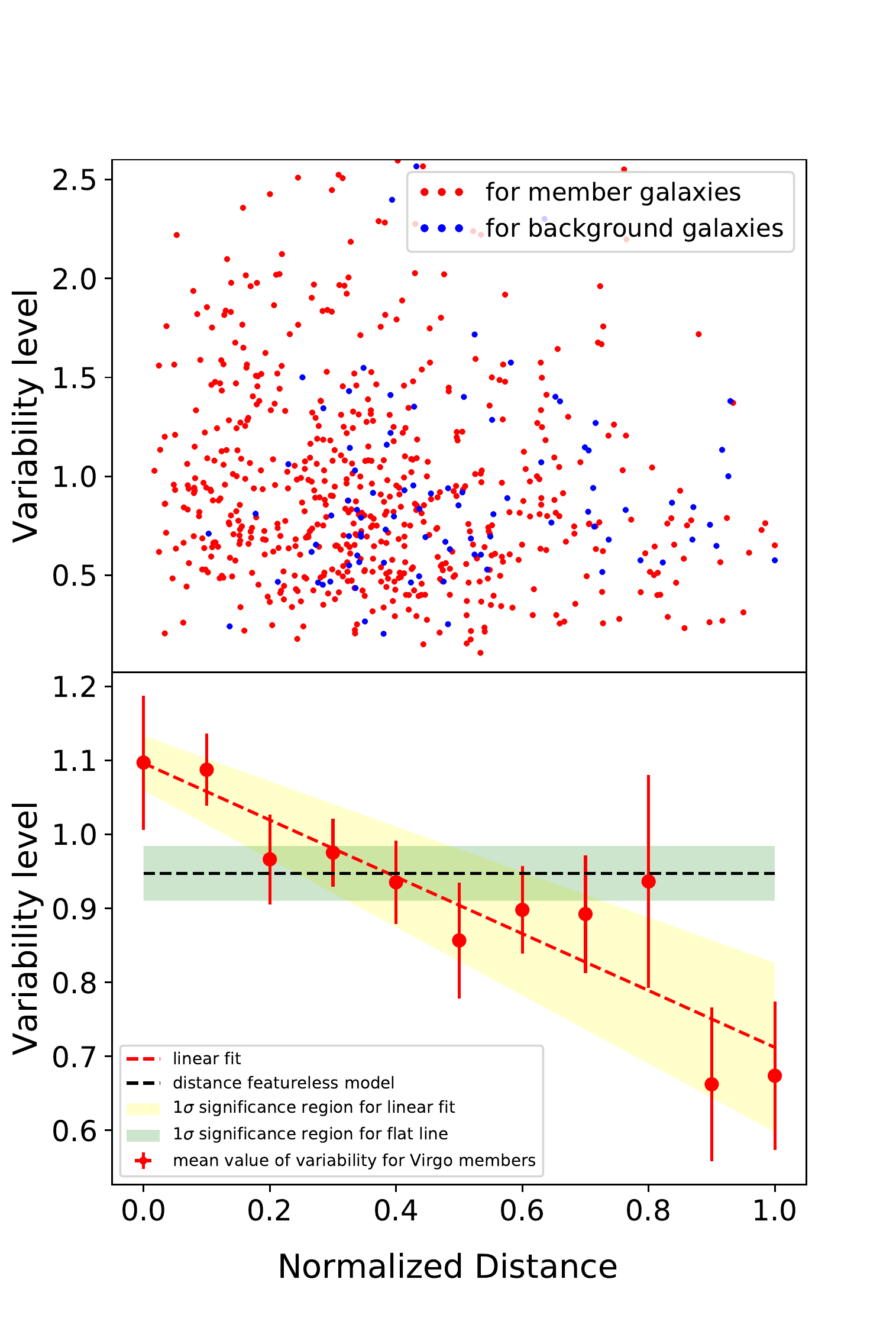}
\caption{Upper panel: Optical variability vs clustercentric distance. The normalized distance is defined as the ratio between galaxy projected distance to the cluster centre and the largest projected distance to the cluster centre among all the galaxy in the sample. The red points give the scattering of Virgo member galaxies. The blue points are scattering of background galaxies for comparison. Bottom panel: Binned galaxy variability for member galaxies. The red line indicates the linear function fit to the data. The black line is the distance independent model fit. The yellow and green regions indicate the 1$\sigma$ confidence region for the fit.}
\end{figure}

\subsection{Comparison with Background Galaxies and Simulation Data} 

We compare the optical variability-distance relation with samples consisting of background galaxies in the vicinity of the Virgo cluster, as well as Monte Carlo simulated galaxies.

Background galaxies in the Virgo cluster neighborhood are also taken from VCC as discussed in sections 2.1 and 2.2. The binned variability shows no significant relation to the distance from the cluster centre as shown in Figure 4. When we fit the data with a linear fit function, the slope is -0.06$\pm$0.10 which shows no significant difference from 0.00. The difference in chi-square between the linear fit (2.88) and the distance-independent fit (2.92) is negligible. The results show no compelling clustercentric distance-dependent relation for background galaxy variability. The background galaxies imply that the clustercentric-dependent variability is a cluster intrinsic property rather than due to some photometry measurement bias in the Virgo Cluster vicinity.

\begin{figure}
\centering
\includegraphics[width=3.8in]{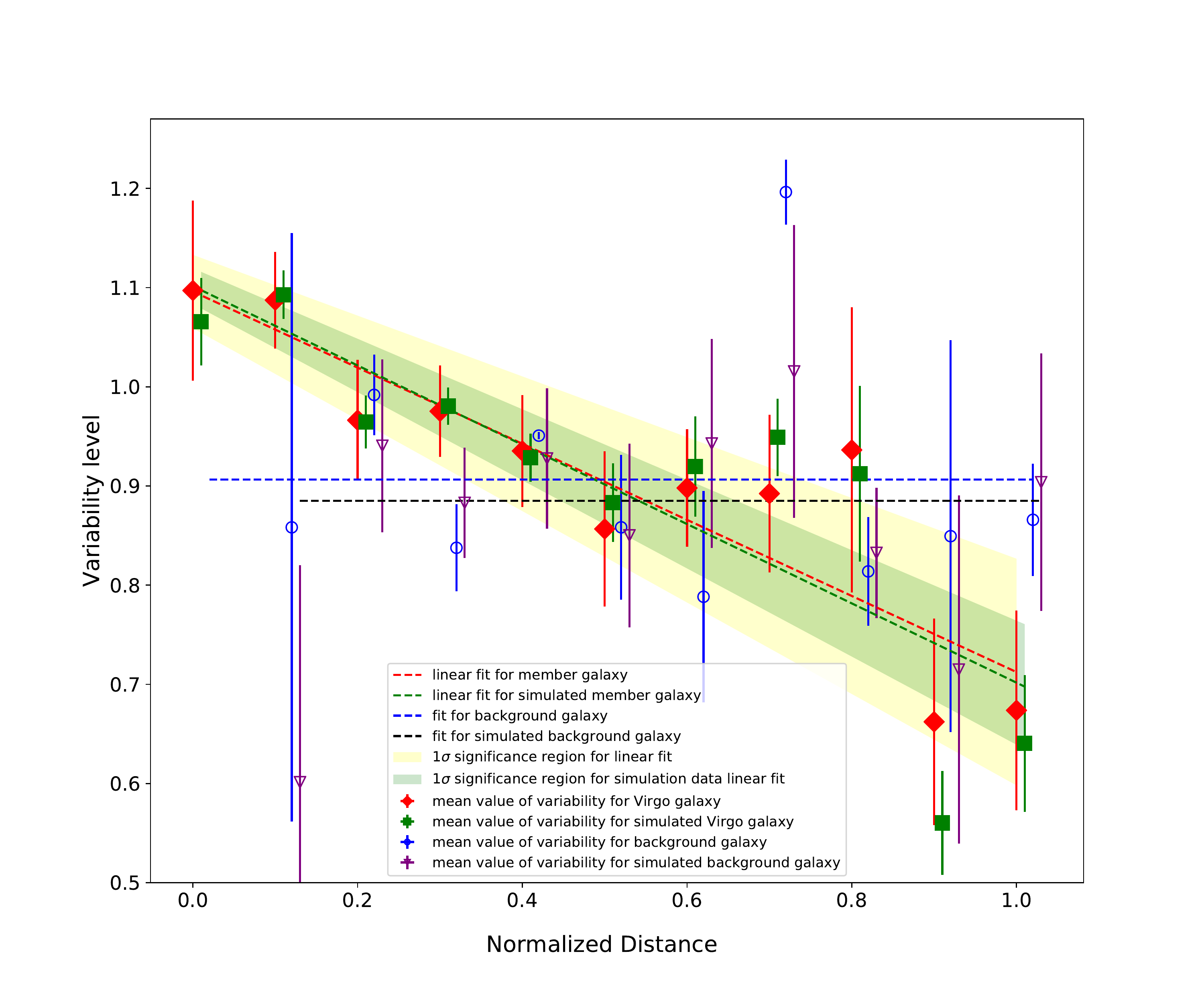}
\caption{The comparison of binned galaxy variability between different samples. The red diamonds indicate the variability of Virgo member galaxies; green squares, the variability of simulated member galaxies; blue circles, the variability of background galaxies; and purple triangles, simulated background galaxies. The green, blue and purple markers are shifted along the x-axis for clarity. The red line is the fit for the member galaxies; green line, simulated member galaxies; blue line, background galaxies; black line, simulated background galaxies.}
\end{figure}

The relation was also investigated with simulated data. The simulation sampled the brightness distribution for every galaxy. We built a Gaussian distribution for every photometry epoch based on the magnitude and its error. This photometry was then sampled using the Gaussian distribution. In this manner, we obtained a new sample of simulated galaxy brightnesses and calculated the mean value of the variability. The procedure was repeated for 1000 times, both for Virgo member galaxies and background galaxies.

The binned galaxy variability from simulated data was then co-added. The median value was treated as the result of the whole simulation (as shown in Figure 4). The standard deviation of the binned variability among all the simulations is set as the error. The variability for member galaxies shows a consistent monotonically decreasing trend comparable to the real data. The results from background galaxies yield no significant relation.

By comparison, the simulation result for background galaxies gives a different data structure than real data. A constant function is a better fit than a linear function with a significant slope for the simulation data. The reduced chi-square is 0.65 for the constant fit and 0.98 for the linear fit. The inconsistent structure for background galaxies, even though both show a distance-independent relation, comes from the limited number of data sets. In contrast, the stable distribution for the member galaxies indicates that the relationship is not due to the data randomly walking within the photometry precision.

\subsection{A Possible Explanation for the Cluster Galaxy Variability Distribution}
Galaxy variability presents a decreasing trend from the cluster centre outwards. The cluster-size feature of galaxy variability is a tracer of the environmental influences on galaxies. A precise description of how galaxies are affected is not clear due to limited observational evidence. There are a number of possible explanations but they are not fully conclusive. For example, more massive galaxies in the cluster centre might present more nuclear activity, according to the correlation between AGN luminosity and host galaxy mass \citep{Fabian2012}. However, the AGN fraction is reported to statistically decrease in the cluster centre \citep{Miller2003,Kauffmann2004} which indicates that the increase of variability towards the cluster centre is specific to the Virgo cluster rather than for clusters in general. Another possibility is that the galaxies may be disturbed with gas provided by galaxy mergers \citep{Hopkins2006}. However, whether a merger is dry or wet differs in observations \citep{Heckman2014}. Here, we present one possibility to explain the cluster scale galaxy variability distribution according to the best of our knowledge.

The Virgo cluster is reported as a cooling flow cluster \citep{Hudson2010} in which group member galaxies show increasing star formation towards the cluster centre \citep{Peterson2003}. The central giant elliptical galaxy M87 is also known to present a high star formation rate due to the accretion of cooling intracluster gas \citep{Fabian1982}.  \citet{Hillel2014} states that some dense cool clumps are not fully heated by the feedback. Some of the cool gas falls in and feeds the central supermassive black hole. If the cooling gas in the Virgo cluster behaves accordingly to the theory by \citet{Hillel2014}, the gas will provide fuel for AGN activity which is consistent with our observations. The explanation fits but more observations, e.g. a comprehensive investigation of the variability of member galaxies for different types of galaxy clusters, would shed light on the underlying physical processes.

\section{Summary and Discussion}
Using PTF time-domain data, we investigate the relationship between galaxy optical variability and clustercentric distance for the Virgo cluster. We define a measure to quantify the galaxy variability taking into account the photometry uncertainty-brightness dependency. The binned galaxy variability for Virgo member galaxies implies a monotonically decreasing variability with increasing clustercentric distance. The variability level is 3.2$\sigma$ higher in the cluster centre than outside of it. From fitting a flat function to the variability data we reject a distance-independent relationship.

The comparison with background galaxies in the Virgo vicinity shows no significant trend implying that the relationship is not due to any photometry bias in the cluster field of view. We also use Monte Carlo simulation for rebuilding the brightness of each galaxy based on its measured photometry and uncertainty. The simulation is repeated for 1000 times both for member and background galaxies. The simulation result implies that the relation for member galaxies is real, and not due to data fluctuation. By contrast, the simulation result for background galaxy oscillates and does not reveal any monotonic function.

Our work shows the environmental influence on member galaxies in a cooling flow cluster. The tracer is from time-domain photometry. The optical variability of AGN has been known for many years, while its application to AGN selection is more recent. To the best of our knowledge, this is the first time that it has been used as a statistical tracer in a cooling flow cluster. The cluster size variability feature can be explained with cooling flow phenomena and theory. Our techniques are general and are suitable for application to other clusters for investigating the influence of the environment upon cluster galaxies. Present and future surveys like the Zwicky Transient Facility (ZTF), and the Large Synoptic Survey Telescope (LSST) will provide more data for this purpose.

\section{acknowledgments}

We wish to thank the referee for their most useful comments which have greatly improved the paper. This work made use of the IPAC database\footnote{\url{https://irsa.ipac.caltech.edu/cgi-bin/Gator/nph-dd}} and PyAstronomy\footnote{https://github.com/sczesla/PyAstronomy} \citep{pya}. We would like to thank Prakash Abhishek, Lee Bomee and Thomas Kupfer for useful discussions. We also thank You-Jun Lu for feedback on our work. Fan Yang, Su-Su Shan and Ji-Feng Liu acknowledge funding from the National Natural Science Foundation of China (NSFC.11988101), the National Science Fund for Distinguished Young Scholars (No.11425313) and the National Key Research and Development Program of China (No.2016YFA0400800). Jun-Qiang Ge acknowledges support from NSFC (No. 11903046), and by the Beijing Municipal Natural Science Foundation under grant No. 1204038.

\bibliographystyle{mnras}
\bibliography{ref}

\bsp	
\label{lastpage}
\end{document}